\documentstyle[12pt]{elsart}
\def\lsim{\mathrel{\rlap {\raise.5ex\hbox{$ < $}}
{\lower.5ex\hbox{$\sim$}}}}

\newcommand{\be}{\begin{equation}}
\newcommand{\ee}{\end{equation}}
\newcommand{\bea}{\begin{eqnarray}}
\newcommand{\nn}{\nonumber}
\newcommand{\eea}{\end{eqnarray}}

\def\gappeq{\mathrel{\rlap {\raise.5ex\hbox{$>$}}
{\lower.5ex\hbox{$\sim$}}}}

\def\lappeq{\mathrel{\rlap{\raise.5ex\hbox{$<$}}
{\lower.5ex\hbox{$\sim$}}}}

\def\beq{\begin{equation}}
\def\eeq{\end{equation}}

\def\Tr{{\rm Tr}\,}
\catcode`\@=11 
\def\coeff#1#2{{\textstyle{#1\over #2}}}

\def\ket#1{\left| #1\right\rangle}
\def\VEV#1{\left\langle #1\right\rangle}

\def\lsim{\mathrel{\mathpalette\@versim<}}
\def\gsim{\mathrel{\mathpalette\@versim>}}
\def\@versim#1#2{\vcenter{\offinterlineskip
    \ialign{$\m@th#1\hfil##\hfil$\crcr#2\crcr\sim\crcr } }}

\def\t1{{\tilde 1}}

\def\to{\rightarrow}

\def\gappeq{\mathrel{\rlap {\raise.5ex\hbox{$>$}}
{\lower.5ex\hbox{$\sim$}}}}

\def\lappeq{\mathrel{\rlap{\raise.5ex\hbox{$<$}}
{\lower.5ex\hbox{$\sim$}}}}

\begin{document}
 
%
%
\begin{flushright}
CERN-TH/99-60 \\
OUTP-99-17P \\

hep-ph/9903386 \\
\end{flushright}

\begin{centering}
\vspace{.2in}
{\large {\bf Comments on CP, T and CPT Violation in Neutral Kaon Decays}}
\\
\vspace{.2in}

 {\bf John Ellis}$^{a}$  and 
{\bf N.E. Mavromatos}$^{a,b}$

\vspace{.4in}

$^{a}$ CERN, Theory Division, CH-1211 Geneva 23, Switzerland. \\
$^{b}$ University of Oxford, Department of Physics, Theoretical Physics,
1 Keble Road, Oxford OX1 3NP, U.K.  \\

\vspace{.3in}

{\it Contribution to the Festschrift for L.B. Okun, \\
to appear in a special issue of Physics Reports \\
eds. V.L. Telegdi and K. Winter}

\vspace{.3in}
 
{\bf Abstract} \\
\vspace{.1in}
\end{centering}
{\small 
We comment on CP, T and CPT violation
in the light of 
interesting new data from the
CPLEAR and KTeV Collaborations on neutral 
kaon decay asymmetries.
Other recent data
from the CPLEAR experiment, constraining possible
violations of CPT and the $\Delta S = \Delta Q$ rule,
exclude the possibility 
that the semileptonic-decay asymmetry $A_{\rm T}$ 
measured by CPLEAR could be solely
due to CPT violation, confirming that 
their data constitute direct evidence for T violation.
The CP-violating asymmetry in $K_L \rightarrow e^-e^+\pi^-\pi^+$
recently measured by the KTeV Collaboration does not 
by itself 
provide
direct evidence for T violation, but we use it to place new
bounds on CPT violation.}


\section{Introduction} 

Ever since the discovery of CP violation in $K^0_L \rightarrow 2 \pi$
decay by Christenson, Cronin, Fitch and Turlay in 1964~\cite{cp}, its
understanding has been a high experimental and theoretical
priority. Until recently, mixing in the $K^0 - {\overline K^0}$
mass matrix was the only known source of CP violation, since it was 
sufficient by itself to explain the observations of CP violation in other
$K^0_{S,L}$ decays, no CP violation was seen in experiments on
$K^{\pm}$, charm or $B$-meson decays, and
searches for electric dipole moments only gave upper limits~\cite{PDG}. 
There has in parallel been active discussion whether the observed
CP violation should be associated with the violation of T or
CPT~\cite{bs}.
Stringent upper limits on CPT violation~\cite{Okun} in the $K^0 -
{\overline K}^0$
system have been given~\cite{CPT}, 
in accord with the common theoretical prejudice based on a
fundamental theorem in quantum field theory~\cite{luders}.
This suggests strongly that T must be violated, but, 
at least until recently, there was no direct observation of T violation.
An indirect demonstration of T violation in neutral kaons,
based on a phenomenological analysis of CP-violating amplitudes,
was made in 1970 using data on the decay 
of long- and short-lived kaons into 
two neutral pions~\cite{schubert}. However,
that analysis assumed unitarity, namely that
kaons disappeared only into the observed states. 

The accumulation of experimental observations of CP and T violation has
accelerated abruptly in the past few months. There have been
two results on $K^0,{\overline K^0}$
decays for which interpretations as direct observations of
T violation have been proposed. One is an
asymmetry in $p{\overline p}$ annihilation,
$p{\overline p} \rightarrow K^-\pi^+K^0~
{\rm or}~K^+\pi^-{\overline K}^0$~\cite{cplearat}, and the other is
a T-odd angular asymmetry in $K^0_L \rightarrow \pi^+ \pi^- e^+ e^-$
decay~\cite{ktev}. 
More recently, a tantalizing hint has been presented
that CP may be violated at a high level in $B^0 \rightarrow J/\psi K_S$
decays~\cite{CDF}. 
Most recent of all, a previous measurement of direct CP violation in
the amplitudes for
$K^0_{S,L} \rightarrow 2 \pi$ decays~\cite{NA31} 
has now been confirmed
by the KTeV Collaboration~\cite{ktev}, 
providing an improved determination of a second
independent CP-violating
experimental number, namely $\epsilon'/\epsilon$,
to test theories and to discriminate between them.

One casualty of this measurement of $\epsilon'/\epsilon$ has
been the superweak theory~\cite{Wolf}, 
according to which all CP violation
should be ascribed to mass mixing in the $K^0 - {\bar K^0}$
system. Still surviving is the Kobayashi-Maskawa model of
weak charged-current mixing within the Standard Model with six 
quarks~\cite{KM}.
Indeed, the new KTeV result arrives
23 years after $\epsilon'/\epsilon$ was first calculated within
the Kobayashi-Maskawa model~\cite{EGN}, 
and it was pointed out that this
would be a (difficult) way to discriminate between this and
the superweak theory, providing (at least part of) the
motivation for this experiment. Coincidentally, the value estimated
there agrees perfectly with the current world average for
$\epsilon'/\epsilon$, although many new diagrams and numerical
improvements have intervened~\cite{buras}. 
The latest theoretical wisdom
about the possible value of $\epsilon'/\epsilon$ within the Standard Model
is consistent with the value measured, at least if the strange-quark
mass is sufficiently small~\cite{sanda}. Thus $\epsilon'/\epsilon$ does not cry
out for any extension of the Standard Model, such as
supersymmetry~\cite{masiero}, though this cannot be excluded. 

It is not the purpose of this article to review in any detail
the potential significance of the $\epsilon'/\epsilon$
measurements, or of the hint of a CP-violating  asymmetry in
$B^0 \rightarrow J/\psi K_S$. Rather, we wish to comment on the suggested
interpretations of the asymmetry in 
$p{\overline p} \rightarrow K^-\pi^+K^0~
{\rm and}~K^+\pi^-{\overline K}^0$~\cite{cplearat}, and of the
T-odd angular asymmetry in $K^0_L \rightarrow \pi^+ \pi^- e^+ e^-$
as possible direct evidence for T violation~\cite{ktev}. 
We argue that the
former can indeed be interpreted in this way, when combined
with other CPLEAR data constraining the possible violation
of the $\Delta S = \Delta Q$ rule and CPT violation in
semileptonic $K^0$
decays~\cite{cplearcpt,cpleary}. We use  the $K^0_L \rightarrow \pi^+
\pi^- e^+ e^-$
decay asymmetry as a novel test of CPT
invariance in decay amplitudes, though one that may not yet be comparable
in power with other tests of CPT.

The layout of this article is as follows: in Section 2 we first
introduce the semileptonic-decay asymmetry recently measured by
CPLEAR, then in Section 3 we introduce a density-matrix description that
includes a treatment of unstable particles as well as allowing
for the possibility of stochastic CPT violation~\cite{ehns,emn,elmn}. 
In Section 4 we
apply this framework to show that the CPLEAR asymmetry cannot be
due to CPT violation, and is indeed a
direct observation of T violation. We also comment whether other
examples of CP violation can be mimicked by CPT violation~\cite{elmn}. 
Then, in
Section 5 we analyze the
decay asymmetry observed by the KTeV collaboration,
arguing that it 
does not have an unambiguous interpretation 
as a direct observation of
T violation. It could not be
due to CPT violation in the mass-mixing
matrix, but could in principle be due to `direct' CPT violation in a decay
amplitude.

\section{The CPLEAR Asymmetry in
$p{\overline p} \rightarrow K^-\pi^+K^0$ and
$K^+\pi^-{\overline K}^0$}

We first recall briefly
the key features of the asymmetry $A_{\rm T}$ observed by CPLEAR,
motivating its interpretation 
as direct evidence for T violation.
The essential idea is to look for a violation of reciprocity
in the rates for $K^0 \rightarrow {\overline K}^0 $ and the
time-reversed reaction ${\overline K}^0 \rightarrow {K}^0 $,
denoted by $P_{K{\overline K}}$ and $P_{{\overline K}K}$,
respectively, as expressed in the asymmetry
\be
   A_{\rm T} \equiv - \frac{P_{K{\overline K}} - P_{{\overline
K}K}}{P_{K{\overline K}} + P_{{\overline K}K}}
\label{at}
\ee
CPLEAR has the unique capability to tag the initial $K^0$
or ${\overline K}^0$ by observing an accompanying $K^{\pm}\pi^{\mp}$
pair in a $p{\overline p}$ annihilation event. However, it is
also necessary to tag the $K^0$ or ${\overline K}^0$ at some later
time, which CPLEAR accomplishes using
semileptonic decays, and constructing the observable
asymmetry~\cite{cplearat}:
\be
  A_l \equiv -\frac{R[\pi^+K^-{\pi^+e^-{\overline \nu}}]- 
R[\pi^-K^+{\pi^-e^+\nu}]}{R[\pi^+K^-{\pi^+e^-{\overline \nu}}]+ 
R[\pi^-K^+{\pi^-e^+\nu}]}
\label{al1}
\ee
where rates are denoted by $R$.
If one assumes the $\Delta S=\Delta Q$ rule, whose 
validity has been confirmed independently
by CPLEAR (see below), then (\ref{al1}) may be re-expressed as
\be
   A_l = \frac{P_{{\overline K} K}(\tau)
BR[K^0 \rightarrow \pi^-e^+\nu] 
- P_{K {\overline K}}(\tau)BR[{\overline K}^0 \rightarrow
\pi^+e^-{\overline \nu}]}
{P_{{\overline K} K}(\tau)
BR[K^0 \rightarrow \pi^-e^+\nu] 
+ P_{K {\overline K}}(\tau)BR[{\overline K}^0 \rightarrow
\pi^+e^-{\overline \nu}]} 
\label{al}
\ee
where decay branching ratios are denoted by $BR$.
In our discussion below, we consider both the cases where the
$\Delta S = \Delta Q$ rule is assumed and where it
is relaxed~\footnote{A recent theoretical discussion using the
$\Delta S = \Delta Q$ rule and CPT invariance
is given in \cite{lola}.}.

If one assumes CPT invariance in the semileptonic-decay
amplitudes,
as was done in the CPLEAR analysis~\cite{cplearat}, 
then $A_l = A_{\rm T}$ and the asymmetry observed by 
CPLEAR can be interpreted as T violation. 
Some doubts about this interpretation have been
expressed~\cite{wolfenstein}, apparently based on concerns
about the inapplicability of the reciprocity arguments of~\cite{kabir}
to unstable particles. We do not believe this to be a problem, since
the analysis of~\cite{kabir} can be extended
consistently to include unstable particles~\cite{elmn,lola,kabir2}.

However, it has also been proposed~\cite{kabir2} that one might
be able to maintain T invariance, 
$P_{{\overline K}K}=P_{K{\overline K}}$, interpreting
the asymmetry observed by CPLEAR instead as
CPT violation in the semileptonic-decay amplitudes~\cite{kabir2}.
This interpretation of the CPLEAR result would be more exciting
than the conventional one in terms of T violation. 
It was suggested in~\cite{kabir2} that this hypothesis of CPT violation
could be tested in the semileptonic decays $K_S \rightarrow \pi l \nu$.
However, 
the hypothesis of~\cite{kabir2} can, in fact, already be
excluded by other published CPLEAR data, as we see below.

\section{Density-Matrix Formalism}

Before discussing this in more detail,
we review the density-matrix formalism~\cite{elmn}, 
which is a convenient 
formalism for treating unstable particles, 
and enables us to present a unified phenomenological analysis 
including also the possibility of stochastic CPT violation
associated with a hypothetical open quantum-mechanical
formalism associated with some approaches to
quantum gravity~\cite{hawk,ehns,marinov,emn}.
In fact, as we recall below, this formalism
has already been used in the Appendix of~\cite{elmn} 
to discard the possibility that CP violation 
in the neutral-kaon system could be `mimicked'
by the CPT-violating mass-matrix parameter $\delta$ within 
conventional quantum mechanics. As we discuss later, this
was possible only if
\be
   {\rm Re} (\delta) \sim \left( 1.75 \pm 0.7 \right) \times 10^{-3} 
\label{delta}
\ee
This analysis is also reviewed briefly below, taking into account 
recent data of the CPLEAR collaboration~\cite{cplearcpt} 
on ${\rm Re}(\delta ) $, which were not available at the time of
writing of~\cite{elmn}, and exclude the possibility (\ref{delta}).

When one considers an unstable-particle system
in isolation, without including its decay channels, 
its time-evolution is non-unitary, so one uses a non-Hermitean
effective Hamiltonian:
$H \ne H^\dagger$. The temporal evolution of the density matrix, $\rho$,  
is given within the conventional quantum-mechanical framework by:

\begin{equation}
\partial _t \rho = -i (H\rho - \rho H^{\dagger})\ .
\label{denmatr}
\end{equation}
In the case of the neutral-kaon system,
the phenomenological Hamiltonian contains the following
Hermitean (mass) and anti-Hermitean (decay)
components:
\begin{equation}
  H = \left( \begin{array}{cc}
 (M + \coeff{1}{2}\delta M) - \coeff{1}{2}i(\Gamma + \coeff{1}{2}
 \delta \Gamma)
&   M_{12}^{*} - \coeff{1}{2}i\Gamma _{12}^{*} \\
           M_{12}  - \coeff{1}{2}i\Gamma _{12}
&    (M - \coeff{1}{2}
    \delta M)-\coeff{1}{2}i(\Gamma
    - \coeff{1}{2}
    \delta \Gamma ) \end{array}\right)\ ,
\label{hmatr}
\end{equation}
in the ($K^0$, ${\bar K}^0$) basis. The ${\delta}M$ and ${\delta}{\Gamma}$
terms violate CPT. Following~\cite{ehns},
we define components of $\rho$ and $H$ by
\begin{equation}
\rho \equiv  \coeff{1}{2}\rho _{\alpha} \sigma _{\alpha}
\qquad ; \qquad H \equiv \coeff{1}{2}h_{\alpha}\sigma _{\alpha}
\qquad ; \qquad \alpha = 0,1,2,3
\label{rhosigma}
\end{equation}
in a Pauli $\sigma$-matrix representation : since the density matrix
must be Hermitean, the ${\rho_{\alpha}}$ are
real, but the $h_{\beta}$ are complex in general. 

We may represent
conventional quantum-mechanical evolution by
$\partial_t\rho_\alpha =H_{{\alpha}{\beta}}{\rho_{\beta}}$, where, in the
($K^0$, ${\bar K}^0$) basis and allowing for the possibility of CPT violation,
\begin{equation}
  H_{\alpha\beta} \equiv \left( \begin{array}{rrrr}
 {\rm Im}h_0 & {\rm Im}h_1 & {\rm Im}h_2 & {\rm Im}h_3 \\
 {\rm Im}h_1 & {\rm Im}h_0 & -{\rm Re}h_3 & {\rm Re}h_2 \\
 {\rm Im}h_2 & {\rm Re}h_3 & {\rm Im}h_0 & -{\rm Re}h_1 \\
 {\rm Im}h_3 & -{\rm Re}h_2 & {\rm Re}h_1 & {\rm Im}h_0 \end{array}\right)\ .
\label{habmatr}
\end{equation}
It is convenient for the rest of our discussion to transform to the
$K_{1,2} = \coeff{1}{\sqrt{2}}(K^0 \mp {\bar K}^0)$
basis, corresponding to $ \sigma_1\leftrightarrow\sigma_3$,
$\sigma_2\leftrightarrow-\sigma_2$, in which $H_{\alpha\beta }$ becomes
\begin{equation}
 H_{\alpha\beta}
 =\left( \begin{array}{cccc}  - \Gamma & -\coeff{1}{2}\delta \Gamma
& -{\rm Im} \Gamma _{12} & -{\rm Re}\Gamma _{12} \\
 - \coeff{1}{2}\delta \Gamma
  & -\Gamma & - 2{\rm Re}M_{12}&  -2{\rm Im} M_{12} \\
 - {\rm Im} \Gamma_{12} &  2{\rm Re}M_{12} & -\Gamma & -\delta M    \\
 -{\rm Re}\Gamma _{12} & -2{\rm Im} M_{12} & \delta M   & -\Gamma
\end{array}\right)\ .
\label{hcomp}
\end{equation}
The corresponding equations of motion for the components of $\rho$ in
the $K_{1,2}$ basis are given in~\cite{elmn}.

The CP-violating mass-mixing parameter ${\epsilon}$ and the CPT-violating
mass-mixing parameter $\delta$ are given by~\footnote{We follow here the 
conventions of~\cite{elmn}, which are related to the notation used
elswhere~\cite{maiani} for CP- and CPT-violating 
parameters by
$\epsilon = -\epsilon_M^*,~\delta=-\Delta^*$, with ${}^*$ denoting 
complex conjugation. Thus the superweak angle $\phi_{sw}$ defined
in~\cite{maiani} is related to the angle $\phi$ in (\ref{eps-delta})
by $\phi = \phi_{sw} -\pi$, so that
${\rm tan}\phi_{sw} = {\rm tan}\phi = 2\Delta m / |\Delta \Gamma|$.}.
\begin{equation}
  \epsilon =\frac{{\rm Im} M_{12}}{\coeff{1}{2}|\Delta\Gamma|+i\Delta m
} = |\epsilon|e^{-i\phi}\
,\qquad
\delta= -\coeff{1}{2}\frac{ -\coeff{1}{2}\delta \Gamma+i\delta M}
{\coeff{1}{2}|\Delta\Gamma|+i\Delta m}\ .
\label{eps-delta}
\end{equation}
One can readily verify~\cite{elmn} 
that ${\rho}$ decays at large $t$ to
\begin{equation}
  \rho \sim e^{-\Gamma _Lt}
 \left( \begin{array}{cc}
 1   &  \epsilon^* + \delta^* \\
 \epsilon + \delta & |\epsilon + \delta |^2 \end{array}\right)\ ,
\label{rhodec}
\end{equation}
which has a vanishing determinant, thus corresponding to a pure long-lived mass
eigenstate $K_L$, whose state vector is
\begin{equation}
\ket{K_L} \propto (1+\epsilon-\delta)\ket{K^0} -
 (1-\epsilon+\delta)\ket{{\bar K}^0}
\label{KL}
\end{equation}
Conversely, in the short-time limit a $K_S$ state is represented by
\begin{equation}
  \rho \sim e^{-\Gamma _St}
 \left( \begin{array}{cc}
 |\epsilon - \delta |^2    &  \epsilon - \delta \\
 \epsilon^* - \delta^* & 1  \end{array}\right)\ ,
\label{rhodecs}
\end{equation}
which also has zero determinant and hence represents a pure state:
\begin{equation}
\ket{K_S} \propto (1+\epsilon+\delta)\ket{K^0} +
 (1-\epsilon-\delta)\ket{{\bar K}^0}
\label{KS}
\end{equation}
Note that the relative signs of the $\delta$
terms have reversed between (\ref{rhodec}) and (\ref{rhodecs}): this is
the signature of mass-matrix CPT violation in the conventional
quantum-mechanical formalism, as seen in the state vectors (\ref{KL})
and (\ref{KS}).

The differential equations for the components of $\rho$ may be
solved in perturbation theory in
$|\epsilon|$ and the new parameters
\begin{equation}
\widehat{\delta M}\equiv{\delta M\over|\Delta\Gamma|}\ ,\qquad
\widehat{\delta\Gamma}\equiv{\delta\Gamma\over|\Delta\Gamma|}\ .
\label{app.2}
\end{equation}
To first order, one finds~\cite{elmn}:
\begin{eqnarray}
\rho_{11}^{(1)}&=&-2|X'||\rho_{12}(0)|\left[
e^{-\Gamma_L t}\cos(\phi-\phi_{X'}-\phi_{12})
-e^{-\Gamma t}\cos(\Delta mt+\phi-\phi_{X'}-\phi_{12})\right]\nonumber \\
\rho_{22}^{(1)}&=&-2|X||\rho_{12}(0)|\left[
e^{-\Gamma_S t}\cos(\phi+\phi_{X}+\phi_{12})
-e^{-\Gamma t}\cos(\Delta mt-\phi-\phi_{X}-\phi_{12})\right]\nonumber \\
\rho_{12}^{(1)}&=&\rho_{11}(0)|X|e^{-i(\phi+\phi_X)}
\left[e^{-\Gamma_L t}-e^{-(\Gamma+i\Delta m)t}\right] \nonumber\\
&~&~~~~~~~~~~~~~~~+\rho_{22}(0)|X'|e^{i(\phi-\phi_{X'})}
\left[e^{-\Gamma_S t}-e^{-(\Gamma+i\Delta m)t}\right]\nonumber\\
&&
\end{eqnarray}
where 
the two complex constants $X$ and $X'$ are defined by:
\begin{eqnarray}
X&=&|\epsilon|+\coeff{1}{2}\cos\phi\,\widehat{\delta\Gamma}
+i\cos\phi\,\widehat{\delta M}\, ,\qquad
\tan\phi_{X}={\cos\phi\,\widehat{\delta M}\over
|\epsilon|+\coeff{1}{2}\cos\phi\,\widehat{\delta\Gamma}}\ ,
\label{X}\nonumber \\
X'&=&|\epsilon|-\coeff{1}{2}\cos\phi\,\widehat{\delta\Gamma}
+i\cos\phi\,\widehat{\delta M}\, ,\qquad
\tan\phi_{X'}={\cos\phi\,\widehat{\delta M}\over
|\epsilon|-\coeff{1}{2}\cos\phi\,\widehat{\delta\Gamma}}\ .
\label{X'}
\end{eqnarray}
The special case that occurs when $\delta M=0$
and $|\epsilon|=0$, namely
\begin{eqnarray}
\delta\Gamma > 0:&&\phi_{X}=0,\quad \phi_{X'}=\pi\; \nonumber \\ 
\delta\Gamma < 0:&&\phi_{X}=\pi,\quad \phi_{X'}=0.
\label{special} 
\end{eqnarray}
will be of particular interest for our purposes.

With the results for $\rho$ through first order, and inserting the appropriate
initial conditions~\cite{elmn} we can immediately write down
expressions for various observables~\cite{elmn} of relevance to CPLEAR.
The values of observables $O_i$ are given in this
density-matrix formalism by expressions of the form~\cite{ehns}
\begin{equation}
     \VEV{O_i} \equiv {\rm Tr}\,[O_i\rho]\ ,
\label{13and1/2}
\end{equation}
where the observables $O_i$ are represented by $2 \times 2$ Hermitean
matrices. Those associated with the decays of neutral kaons to $2\pi$,
$3\pi$ and $\pi l \nu$ final states are of particular interest to us.
If one assumes the $\Delta S = \Delta Q$ rule, their expressions in the
$K_{1,2}$ basis are
\begin{eqnarray}
 O_{2\pi} &\propto& \left( \begin{array}{cc} 0 & 0 \\
0 & 1 \end{array} \right)\ ,\qquad  O_{3\pi} \propto
\left( \begin{array}{cc} 1 & 0 \nonumber \\
0 & 0 \end{array} \right)\ , \label{2pi-obs} \\
O_{\pi^-l^+\nu} &\propto& \left( \begin{array}{cc}
1 & 1 \\1 & 1 \end{array} \right)\ ,\qquad
O_{\pi^+l^- \bar\nu} \propto \left( \begin{array}{rr}
1 & -1 \\
-1 & 1 \end{array} \right)\ .
\label{semi-obs}
\end{eqnarray}
which constitute a complete Hermitean set. We consider later the
possible relaxation of the $\Delta S = \Delta Q$ rule, and also
the possibility of
direct CPT violation in the
observables (\ref{semi-obs}), which would give them different
normalizations. 
The small experimental value of $\epsilon '/\epsilon$ would be
taken into account by different magnitudes for $O_{2\pi}$ in the
charged and neutral modes, but we can neglect this refinement for
our purposes.

In this formalism, pure $K^0$ or ${\bar K}^0$ states, such those
provided as initial conditions in the CPLEAR experiment,
are described by the following density matrices:
\begin{equation}
\rho _{K^0} =\coeff{1}{2}\left( \begin{array}{cc}
1 &1 \\1 & 1 \end{array} \right)\ , \qquad
\rho _{{\bar K}^0} =\coeff{1}{2}\left( \begin{array}{rr}
1 & -1 \\-1 & 1 \end{array} \right)\ .
\label{rhos}
\end{equation}
We note the similarity of the above density matrices (\ref{rhos})
to the 
representations (\ref{semi-obs}) of the semileptonic decay observables,
which reflects
the strange-quark contents of the neutral kaons 
and our assumption of the validity of the $\Delta S = \Delta Q$ rule:
$K^0 \ni {\bar s}
\rightarrow {\bar u} l^+ {\nu} , {\bar K}^0 \ni s \rightarrow  u l^-
\bar\nu$.

\section{Interpretation of the CPLEAR Asymmetry}

In the CPLEAR experiment~\cite{cplearat}, the
generic quantities measured are asymmetries of 
decays from an initially pure $K^0$ beam as compared to 
the corresponding decays from an initially pure ${\bar K}^0$ beam:
\begin{equation}
    A (t) = \frac{
    R({\bar K}^0_{t=0} \rightarrow
{\bar f} ) -
    R(K^0_{t=0} \rightarrow
f ) }
{ R({\bar K}^0_{t=0} \rightarrow
{\bar f} ) +
    R(K^0_{t=0} \rightarrow
f ) }\ ,
\label{asym}
\end{equation}
where $R(K^0_{t=0}\rightarrow f)\equiv \Tr[O_{f}\rho (t)]$ denotes the
decay rate
into a final state $f$, starting from a pure $ K^0$ at $t=0$: $\rho(t=0)$
is given by the first matrix in (\ref{rhos}), and correspondingly
$R({\bar K}^0 \rightarrow {\bar f}) \equiv \Tr [O_{\bar f} {\bar \rho}(t)]$
denotes the decay rate into the conjugate state ${\bar f}$, 
starting from a pure ${\bar K}^0$ at $t=0$: ${\bar \rho}(t=0)$
is given by the second matrix in (\ref{rhos}).
Several relevant asymmetries were defined in~\cite{elmn},
including $A_{\rm T}$ (already introduced above), $A_{{\rm CPT}}$, $A_{2\pi}$ and
$A_{3\pi}$. We discuss below their
possible roles in discriminating between
CP- and CPT-violating effects, in particular when CPT violation
is invoked so as to mimic CP violation whilst preserving T
invariance~\cite{kabir2}.  

In order to parametrize
a possible CPT-violating difference in semileptonic-decay amplitudes
as postulated there, we define $y$:
\be
<\pi^+e^-{\overline \nu}|{\cal T}|{\overline K}^0> \equiv (1 +
y)<\pi^-e^+\nu|{\cal T}|K^0> 
\label{y}
\ee
and we assume that $y$ is {\it real}, which is justified if the amplitude
is T invariant~\cite{maiani}. We assume this here because the
purpose of this analysis is to test the hypothesis~\cite{kabir2}
that the CPLEAR
asymmetry can be reproduced by CPT violation alone, retaining 
T invariance  
in the mixing: $P_{K{\overline K}} = 
P_{{\overline K}K}$.~\footnote{For clarity and completeness, we note 
the following relation between the quantity $y$ defined above and the
quantity y defined in~\cite{cplearat}: $ y = 2{\rm y} \equiv -2b/a  $
to lowest order in $y$ and for real $a,b$~\cite{maiani}: 
$<\pi^+e^-{\overline \nu}|{\cal T}|{\overline K}^0> = a^*-b^*,~
<\pi^-e^+\nu|{\cal T}|K^0> = a+b$.}
Another important point~\cite{cplearat} is the independence of the 
asymmetry $A_{\rm T}$ measured at late times of any possible 
violation of $\Delta S = \Delta Q$ rule. 
As seen from~\cite{cplearat}, violations of this rule may be
taken into account simply by introducing the combination
\be
{\rm {\tilde y}} \equiv y + 2{\rm Re}(x_{-})
\label{ytilda}
\ee
where, in the notation of~\cite{maiani,cplearat},
 $x_-$ parametrizes violations of the $\Delta S =
\Delta Q$ rule:
\be
<\pi^+e^-{\overline \nu}|{\cal T}|K^0> \equiv c + d, \qquad  
<\pi^-e^+\nu|{\cal T}|{\overline K}^0> \equiv c^*-d^*
\label{cd}
\ee
and $x \equiv  (c^* - d^*)/(a + b) $,
${\overline x}^* \equiv (c + d)/(a^* - b^*)$, and 
$x_{\pm} \equiv  (x \pm {\overline x})/2$.  
Again, the hypothesis of T invariance implies the reality
of $x, {\overline x}, x_{\pm}$~\cite{maiani}. 
If one considers violations of $\Delta S = \Delta Q$ rule,
one should take appropriate account of the additional decay modes
(\ref{cd}) 
in $A_l$ (\ref{al}).

In  the density-matrix formalism, $y \ne 0$ corresponds 
to a difference in normalization
between the semileptonic observables $O_{\pi l\nu}$ 
introduced in (\ref{semi-obs}). The
analysis of~\cite{kabir2},
extended in the above straightforward way to take into account 
of possible violations of the $\Delta S = \Delta Q$ rule, 
shows that, if one imposes
reciprocity, 
then 
\be
A_l \simeq  -{\rm {\tilde y}}
\label{relatey}
\ee
to lowest order in ${\rm {\tilde y}}$.  

To make contact with the experimental measurement of the 
CPLEAR collaboration, one should take into account the different 
normalizations of the $K^0$ and ${\overline K}^0$ 
fluxes at the production point.
Because of this effect,
the measured asymmetry~\cite{cplearat} becomes:
\be
   A_{\rm T}^{exp} = A_l -{\rm {\tilde y}} 
\label{expat}
\ee
The measured~\cite{cplearat} value of this asymmetry is:
\be
   A_{\rm T}^{exp} \simeq \left(6.6 \pm 1.3_{stat} \right)\times 10^{-3} 
\label{atexp2}
\ee
If this experimental result
were to be interpreted as expressing CPT violation but T invariance,
then ${\rm {\tilde y}}$ should have the value:
\be
  {\rm {\tilde   y}}=-\left(3.3 \pm 0.7 \right)\times 10^{-3}
\label{ykabir}
\ee
Such a scenario is excluded 
by
the current CPLEAR value of ${\rm {\tilde y}}$~\cite{cpleary}. 
The late-time asymmetry measured by CPLEAR 
can be expressed as~\cite{cplearat}:
\be
 A_{\rm T}^{exp} \simeq 
4{\rm Re}(\epsilon ) - 2{\rm Re}({\rm {\tilde y}}) 
\label{atexp}
\ee
This enables a stringent upper limit to be placed~\cite{cpleary}:
\be
\frac{1}{2} {\tilde {\rm y}} =
\frac{1}{2}{\rm Re}(y) + {\rm Re}(x_{-}) 
= \left( 0.2 \pm 0.3_{stat}\right) \times 
10^{-3} 
\label{yx}
\ee
Therefore, 
the CPT-violating but T-conserving hypothesis is conclusively
excluded independently of any assumption about the 
validity of the $\Delta S = \Delta Q$ rule.

As a side-remark, we comment on the effect of $y$ 
on the CPT-violating width difference $\delta \Gamma$,
assuming the validity of the $\Delta S = \Delta Q$ rule ($x_{-}=0)$,
which is supported by~\cite{cpleary}.
Using
$\Gamma _{sl}^L =0.39 \times \frac{1}{\tau_L} \simeq 8 \times 10^{-18}
~{\rm GeV}$ and the value (\ref{ykabir}) for $y$, and neglecting any
possible other CPT-violating differences in decay rates, we find
\be
   \delta \Gamma \simeq 1.06 \times 10^{-19}~{\rm GeV} 
\label{dg}
\ee
which makes the following contribution to 
$Re({\delta})$:
\be
  2 {\rm Re}(\delta ) = 
\frac{\delta \Gamma |\Delta \Gamma| }{|\Delta \Gamma |^2 
+ 4|\Delta m|^2} = \frac{\delta \Gamma {\rm cos}^2\phi }{|\Delta \Gamma | }
\simeq 6.8 \times 10^{-6} 
\label{red}
\ee
where we have used $\phi \simeq 43.49^o$ mod $\pi$. This contribution is
far below the present experimental sensitivity discussed below.

Next, we comment on the possibility that
what we usually regard as CP violation in the mass matrix
is actually due to CPT violation.
In such a case, one would have to
set $|\epsilon|\to0$ and make the following choices 
for the CPT-violating mixing parameters
\begin{equation}
{\rm mimic\ CP\ violation:}\qquad\qquad\delta M=0,\qquad
\widehat{\delta\Gamma}\to{2|\epsilon|\over\cos\phi}\ ,
\label{mimic}
\end{equation}
On account of (\ref{special}), then,  
the observable $A_{\rm T}$ would have the following
time-independent first-order expression:
\begin{equation}
A_{\rm T}=
2|X'|\cos(\phi-\phi_{X'})+2|X|\cos(\phi+\phi_X)=4|\epsilon|\cos\phi\ ,
\label{AQM_T}
\end{equation}
which is identical to the conventional case of CPT symmetry.
However, this is not the case for all observables, for instance
the $A_{{\rm CPT}}$ asymmetry, defined
by setting $f= \pi^-e^+\nu $, ${\overline f}= \pi^+e^-{\overline \nu} $
in (\ref{asym}). 
In
particular, 
one has the following asymptotic formula for $A_{\rm CPT}$:
\begin{equation}
A_{\rm CPT}\to4\sin\phi\cos\phi\,\widehat{\delta M}
-2\cos^2\phi\,\widehat{\delta\Gamma}\ .
\label{asymacpt}
\end{equation}
which would yield the following asymptotic
prediction under the
``mimic" assumption (\ref{mimic}):
\begin{equation}
A_{\rm CPT}\to-4|\epsilon|\cos\phi \ ,
\label{compare}
\end{equation}
to be contrasted with the standard result that $A_{\rm CPT}=0$
in the absence of CPT violation.

For comparison with experimental data of CPLEAR,
it is useful to express the conventional CPT-violating 
parameter $\delta$ (\ref{eps-delta}) in terms of 
$\widehat{\delta \Gamma}$:
\be
{\rm Re}(\delta )=\frac{1}{2}\widehat{\delta \Gamma} 
{\rm cos}^2\phi = |\epsilon|{\rm cos}\phi >0,
\;\;
{\rm Im}(\delta ) = -\frac{1}{2}\widehat{\delta \Gamma} 
{\rm sin}\phi{\rm cos}\phi
\label{defdelta}
\ee
The experimental asymmetry $A_{\rm T}^{exp}$ (\ref{expat}), then, 
would be obtained 
upon the identification of $A_{\rm T}$ in (\ref{AQM_T}) with $A_l$,    
\be
   A_{\rm T}^{exp} = 4{\rm Re}(\delta) -{\rm {\tilde y}} 
\label{atredel}
\ee
Notice that in principle such a situation 
is consistent with the experimental 
data, given that the combination $A_{\rm T}^{exp}+{\tilde {\rm y}} =4{\rm Re}(\delta) >0$. 
Taking into account (\ref{atexp2}), (\ref{yx}) and (\ref{atredel})
we observe that  
the mimic requirement would imply
\be 
{\rm Re}(\delta )_{mimic} \sim \left(1.75 \pm 0.7 \right) \times 10^{-3}\,
\label{mimicdelta}
\ee
However, the CPLEAR Collaboration has measured~\cite{cplearcpt}
${\rm Re}(\delta) $ using
the asymptotic value of the asymmetry $A_\delta $:
\be
A_\delta \equiv \frac{{\overline R}_+ - R_-(1 + 4{\rm Re}\epsilon _L)}
{{\overline R}_+ + R_-(1 + 4{\rm Re}\epsilon _L)}    
+ \frac{{\overline R}_- - R_+(1 + 4{\rm Re}\epsilon _L)}
{{\overline R}_- + R_+(1 + 4{\rm Re}\epsilon _L)}    
\label{deltaas}
\ee
which asymptotes at large times to $-8{\rm Re}(\delta)$,
independently of any assumption on the $\Delta S = \Delta Q$ rule:
\be
{\rm Re}(\delta ) \simeq 
\left( 3.0 \pm 3.3_{stat} \pm 0.6_{syst}\right) \times 10^{-4} 
\label{cpl}
\ee
in apparent conflict with (\ref{mimicdelta}).

The fact that the CP violation seen in the
mass matrix cannot be mimicked by CPT violation~\cite{Okun} has been known
for a long time. The possible magnitude of CPT violation is
constrained in particular by the consistency between $\phi_{+-}$ and
the superweak phase $\phi_{sw}$.
However, it is possible to mimic CP violation
in any particular observable by a suitable choice of $\delta$.
For example, as was shown in~\cite{elmn}, the
standard superweak result for $A_{2\pi}$ 
may be reproduced by setting $|\epsilon|\to0$
and using (\ref{mimic}),
which give $|X|\to|\epsilon|$ and $\phi_X=0$.
The standard CP-violating result for $A_{3\pi}$
may also obtained with
the choices (\ref{mimic})~\cite{elmn}, which give $|X'|\to|\epsilon|$
and $\phi_{X'}=\pi$, since $\tan(\phi-\pi)=\tan\phi$.
But, as already emphasized, the dynamical equations determining the
density matrix prevent all observables from being mimicked in this way:
this is what we found above with the $A_{{\rm CPT}}$ observable (\ref{compare}),
to be contrasted with the standard result $A_{{\rm CPT}}=0$. Moreover,
as mentioned above, the mimic hypothesis is excluded by the 
recent CPLEAR result (\ref{cpl}). 

It was also pointed out previously~\cite{elmn,hp}
that deviations from conventional closed-system quantum
mechanics of the type discussed in~\cite{ehns}, which lead to 
stochastic CPT violation, also cannot account for the 
CP violation observed in the neutral kaon system. We remind the reader
that generic possible deviations from closed-system quantum-mechanical
evolution in the neutral kaon system - which might arise
from quantum gravity or other stochastic forces - may be described
by the three real parameters $\alpha,\beta,\gamma$ of~\cite{ehns},
if one assumes energy conservation and dominance by $\Delta S = 0$
stochastic effects. These parameters lead to entropy growth,
corresponding to the appearance of an arrow of time and violation
of CPT~\cite{PW}, as has sometimes been suggested in the context
of a quantum theory of gravity. However, this CPT violation cannot be
cast in the conventional quantum-mechanical form discussed above.
The most stringent bounds on 
the stochastic CPT-violating parameters
$\alpha,\beta,\gamma$ have been placed by the CPLEAR
collaboration~\cite{cplearqmv}. They
are not far from the characteristic magnitude
${\cal O}(M_K^2/M_P)$, where the Planck mass $M_P \simeq 10^{19}~{\rm
GeV}$, near the scale at which such effects might first set in~\cite{emn}
if they are due to 
quantum-gravitational effects.

\section{The KTeV Asymmetry in $K_L \rightarrow e^-e^+\pi^-\pi^+$ and its
Interpretation}

Subsequent to the CPLEAR analysis,
the KTeV Collaboration has reported~\cite{ktev} a novel
measurement of a T-odd asymmetry
in the decay of $K_L \rightarrow e^-e^+\pi^-\pi^+$.
Since incoming and outgoing states are {\it not} exchanged in the
KTeV experiment, unlike the CPLEAR measurement comparing
${\overline K}^0 \rightarrow K^0$ and $K^0 \rightarrow {\overline K}^0$
transitions, it cannot provide direct evidence for
T violation. However, it is interesting to discuss the
information this measurement may provide about CP, T and CPT
symmetry.

This decay has previously been analyzed theoretically 
in~\cite{sehgal}, assuming CPT symmetry. The decay
amplitude was decomposed as:
\be
{\cal M}(K_L \rightarrow \pi^+\pi^-e^+e^-)={\cal M}_{Br} + {\cal M}_{M1}
+ {\cal M}_{E1} + {\cal M}_{SD}^{V,A} + {\cal M}_{CR} 
\label{ampl}
\ee
and the various parts of the amplitude (\ref{ampl}) have the following
interpretations:

\noindent
- ${\cal M}_{Br}$:  Amplitude for the Bremsstrahlung process
related to the standard CP-violating $K_L \rightarrow 2 \pi$
amplitude, violating $CP$ just like the conventional $\epsilon$ parameter. 
This amplitude is proportional to a coupling constant~\cite{sehgal} 
\be 
    g_{Br} = \eta_{+-} e^{i\delta _0(M_K^2)}
\label{gbr}
\ee
where $\eta_{+-}$ is the conventional CP-violating parameter,  
whose phase $\phi_{+-}$ is that of $K_L \rightarrow \pi^+\pi^-$:
$\delta _0(M_K^2)$ is the relevant $I = 0$ $\pi^+\pi^-$ phase shift.

\noindent
- ${\cal M}_{M1}$: The magnetic-dipole contribution to the 
amplitude, which is CP-conserving. The corresponding coupling constant 
has a non-trivial phase~\cite{sehgal}:
\be 
    g_{M1} = i |g_{M1}| e^{i{\delta}_1 (m_{\pi\pi}) + \delta\varphi}
\label{moment}
\ee
where  ${\delta}_1$ is the $\pi\pi$ $P$-wave
phase shift. The amplitude is invariant under CPT if
$\delta\varphi = 0$, leaving the prefactor $i$
as a consequence of CPT invariance. The estimate $|g_{M1}| = 0.76$
is given in~\cite{sehgal}.

\noindent
- ${\cal M}_{E1}$: This denotes the electric-dipole contribution.
It is CP-conserving, and its coupling constant $g_{E1}$ 
has been computed in \cite{sehgal}.   
Its phase is related to that of $g_{M1}$ via
${\rm arg}\left(g_{E1}/g_{M1}\right)\simeq \phi_{+-}$.

\noindent
- ${\cal M}_{SD}^{V,A}$: This is the 
contribution originating in the 
short-distance Hamiltonian describing the transition 
$s{\overline d} \rightarrow e^+e^-$. 
Its coupling constant has been calculated in the
Standard Model~\cite{sehgal}, with the result:
\be
    g_{SD} = i(5 \times 10^{-4})
\sqrt{2}\frac{M_K}{f_\pi} e^{i\delta _1(m_{\pi\pi})}
\label{sd}
\ee
where $f_\pi$ is the pion decay constant. 
One could in principle introduce CPT violation 
into this amplitude by allowing
$A({\overline K}^0 \rightarrow \pi^+\pi^-e^+e^-) \ne
A(K^0 \rightarrow \pi^+\pi^-e^+e^-)$. As seen in (\ref{sd}), these
amplitudes may be related to $M_{\overline K}$ and
$M_K$ respectively, which could be different if
CPT is violated.

\noindent 
- ${\cal M}_{CR}$: This denotes the 
CP-conserving contribution due to  a finite 
charge radius of the $K^0$. Its coupling $g_P$    
has the phase of $K_S \rightarrow \pi^+\pi^-$. 

The KTeV~\cite{ktev} Collaboration's
measurement is of a
CP-violating asymmetry ${\cal A}$ in
the angle $\Phi$ between the vectors
normal to the $e^-e^+$  and $\pi^+\pi^-$ planes~\cite{sehgal},
which is related to the particle momenta by:
\be
   {\rm sin \Phi}{\rm cos \Phi}=\eta_l \times \eta_\pi . \left(
\frac{p_+ +  p_-}{|p_+ +  p_-|} \right).   
\left( \eta_l . \eta_\pi \right) 
\label{asymm}
\ee
where the unit vectors 
$\eta_{l,\pi}$ are defined as 
$\eta _l \equiv {k_+ \times k_-}/{|k_+ \times k_-|}$
and $\eta _\pi \equiv \frac{p_+ \times p_-}{|p_+ \times p_-|}$,
with $k_{\pm}$ the lepton momenta and $p_{\pm}$ the pion momenta. 
The observable is a CP asymmetry ${\cal A}$ 
of the process, which we shall discuss below. 
The $\Phi$ distribution
$d\Gamma/d\Phi$ may be written in the following generic
form~\cite{sehgal}:
\be
 \frac{d\Gamma}{d\Phi} = \Gamma _1 {\rm cos}^2\Phi + 
\Gamma _2 {\rm sin}^2\Phi + \Gamma _3 {\rm cos}\Phi {\rm sin}\Phi 
\label{formula}
\ee
where the last term changes sign under the CP 
transformation and is T-odd, i.e., it changes sign when the
particle momenta are reversed. However, it clearly does not
involve switching `in' and `out' states, and so is not a
direct probe of T violation~\footnote{It
is generally agreed that final-state electromagnetic interactions can be
neglected for present purposes. The KTeV collaboration has recently
reported~\cite{moriond} a null asymmetry in the angle between the $\pi^+
\pi^-$ and
$e^+ e^-$ planes in the Dalitz decay $K_L \rightarrow \pi^+ \pi^- (\pi^0
\rightarrow e^+ e^- \gamma)$. This provides a nice check on the
experimental technique, but does not test directly the structure of
the final-state interactions, since the $\pi^0$ decays outside the
Coulomb fields of the $\pi^+ \pi^-$ pair.}.

A detailed functional form for $\Gamma _3$ is 
given in \cite{sehgal}. Following the above discussion of the various
terms in the decay amplitude (\ref{ampl}), this term is
interpreted~\cite{sehgal} in terms of the dominant Bremsstrahlung, 
magnetic-dipole and electric-dipole contributions.
For our purposes, it is sufficient to
note that it involves the coupling constant combinations
${\rm Re}(g_{M1}g^*_{BR})$ and ${\rm Re}(g_{M1}g^*_{E1})$, which 
involve amplitudes with different CP properties, and hence violate
CP manifestly. It depends in particular on
the phase $\phi_{+-}$ 
of the conventional CP-violating $K_L \rightarrow \pi^+ \pi^-$
decay amplitude, via the $K_1$ admixture in the
$K_L$ wave function, which enters in
the M1 amplitude for
$K_L \rightarrow \pi^+\pi^-\gamma$.
The following is the generic structure of the integrated asymmetry
measured by KTeV~\cite{sehgal}:
\be
{\cal A} = \frac{\int _0^{\pi/2} \frac{d\Gamma}{d\Phi}d\Phi - 
\int_{\pi/2}^{\pi} \frac{d\Gamma}{d\Phi}d\Phi}
{\int _0^{\pi/2} \frac{d\Gamma}{d\Phi}d\Phi +
\int_{\pi/2}^{\pi} \frac{d\Gamma}{d\Phi}d\Phi}
\simeq 
{\cal A}_1~{\rm cos}\Theta_1 + 
{\cal A}_2~{\rm cos}\Theta_2\left|\frac{g_{E1}}{g_{M1}}\right| \nn \\
\label{asymmetry}
\ee
where
\be
\Theta_1 \equiv \phi_{+-} + \delta _0 - {\overline \delta}_1 
- \frac{\pi}{2} - {\overline {\delta \varphi}}~{\rm mod}\pi, \qquad   
\Theta_2 \equiv \phi_{+-} - \frac{\pi}{2} - 
{\overline {\delta \varphi}}~{\rm mod}\pi
\label{Thetas}
\ee
and ${\overline \delta}_1$ and ${\overline {\delta \varphi}}$ are
averages of
the $\pi\pi$ $P$-wave 
phase shift and ${\delta \varphi}$, respectively, in the region $m_{2\pi}
< m_K$. Numerical estimates of the
quantities ${\cal A}_{1,2}$ in terms of the different couplings in
(\ref{ampl}) were given in~\cite{sehgal}:
\be
{\cal A}_1 \simeq 0.15, \; {\cal A}_2 \simeq 0.38, 
\label{Avalues}
\ee
leading to
the following prediction for ${\cal A}$: 
\be
{\cal A} \simeq 
0.15 {\rm sin}\left[ \phi_{+-} + \delta_0(m_K^2)
- {\overline \delta}_1 \right]
\label{asymmpred}
\ee
if the CPT-violating phase ${\overline {\delta \varphi}} = 0$.
Using the experimental values $\delta _0 \simeq 40^o$,
${\overline \delta}_1 
\simeq 10^o$ and $\phi_{+-} \simeq 43^o$, (\ref{asymmpred}) becomes
\be
{\cal A} \simeq 0.14 
\label{asymmpredCPT}
\ee
As already mentioned, the experimental value
\be
  {\cal A}_{exp} = \left( 13.5 \pm (2.5) _{stat} \pm (3.0)_{syst} \right)
\%
\label{expasym}
\ee
agrees very well with the theoretical prediction (\ref{asymmpredCPT})
obtained assuming the CPT-violating phase ${\overline {\delta \varphi}} =
0$.

We now analyze how well this
measurement tests CPT, and assess how this test compares
with other tests.
Consider first the Bremsstrahlung contribution: as
mentioned above, the coupling $g_{br}$ has a phase $\phi_{+-}$. 
In principle, CPT violation in the neutral-kaon mass matrix
could shift this phase away from its superweak value $\phi$
by an amount $\delta \phi$:
\be
    |m_{K^0} - m_{{\overline K}^0} | 
\simeq 2\Delta m\frac{|\eta_{+-}|}{sin\phi} |\delta \phi| 
\label{swph}
\ee
where (as always) we neglect effects that are ${\cal O}(\epsilon ') $, and
we recall that
$|\eta_{+-}| \simeq |\epsilon |/{\rm cos}\delta \phi \simeq |\epsilon |$.  
The best limit on such a mass difference is now provided by
the CPLEAR experiment~\cite{CPT}:
\be
  |m_{K^0} - m_{{\overline K}^0} | \le 3.5 \times 10^{-19}~ 
{\rm GeV}~(95\%~{\rm C.L.}),
\label{dmformula}
\ee
The limit (\ref{dmformula}) determines $|\delta \phi| \lsim 0.86^o$,
whereas a combination of previous data from the NA31, E731
and E773 Collaboration yields~\cite{elmn}   
$\delta \phi \lsim \left( -0.75 \pm 0.79\right)^o$.   
Such a phase change $|\delta \phi|$ would change $\cal A$ by an amount
$|\delta {\cal A}|
\lappeq 10^{-3}$, far smaller than the experimental error in
(\ref{expasym}), and also much smaller than the likely
theoretical uncertainties.

We consider next the magnetic-dipole contribution, with the possible
incorporation of a CPT-violating phase ${\overline {\delta \varphi}}$
(\ref{moment}).
To first order in ${\overline {\delta \varphi}}$, the corresponding
change in ${\cal A}$ is:
\be
  \delta {\cal A} \simeq 
\left(0.15 {\rm sin}\Theta_1^{(0)} + 
0.38 {\rm sin}\Theta_2^{(0)}\left|\frac{g_{E1}}{g_{M1}}\right| \right)
{\overline {\delta \varphi }}
\label{dev}
\ee
with $\Theta_{1,2}^{(0)}$
evaluated using (\ref{Thetas}) and assuming ${\overline {\delta \varphi}}
= 0$. 
However, this small-angle approximation is not justified, so we use the
full
expression (\ref{asymmetry}) for ${\cal A}$, and interpret
the experimental value (\ref{expasym}) as implying that
${\cal A} \gappeq 0.096$ 
at the one-standard-deviation level, corresponding to
\be
 0.14{\rm cos}{\overline {\delta \varphi}} - 0.04{\rm sin}{\overline {\delta \varphi}} \ge 0.096 
\label{untitled}
\ee
which leads to  
\be
- 70^o \lappeq {\overline {\delta \varphi}} \lappeq + 40^o
\label{varphilimit}
\ee
for the allowed range of this CPT-violating parameter, where we have used
the estimate~\cite{sehgal} $|{g_{E1}}/{g_{M1}}| \simeq 0.05$, and
not made any allowance for theoretical uncertainties.
 
The range (\ref{varphilimit}) is clearly much wider than
the corresponding scope for a CPT-violating
contribution $\delta \phi$ to the phase $\phi_{+-}$ of $\eta_{+-}$,
and the range would be larger still if we expanded
the allowed range of ${\overline {\delta \varphi}}$ to the 95\% C.L.
limits. We also note in passing that the magnitude of 
the short-distance contribution (\ref{sd})
is so small that no interesting limit on direct CPT violation
in it can be obtained.

We now address the question whether all the KTeV asymmetry could be
due to CPT violation. This would occur if
\be
{\cal A}_1~{\rm cos}\Theta_1^{(0)} +
{\cal A}_2~{\rm cos}\Theta_2^{(0)}\left|\frac{g_{E1}}{g_{M1}}\right| = 0
\label{CPvanish}
\ee
This possibility is disfavoured by the theoretical estimates of
${\cal A}_{1,2}$, but cannot be logically excluded. If (\ref{CPvanish})
were to hold, the KTeV asymmetry could be written in the form
\be
{\cal A} \simeq  {\cal A}_1{\rm sin} {\overline {\delta \varphi}} \left[
{\rm sin}\Theta_1^{(0)} -
{\rm cos}\Theta _1^{(0)}~{\rm tan}\Theta_2^{(0)}
\right]
\label{whatif}
\ee
in which case the experimental value (\ref{expasym}),
at the one-standard-deviation level,
would be reproduced if 
\be
0.13 \lappeq {\cal A}_1 {\rm sin}{\overline {\delta \varphi}} \lappeq 0.22
\label{finalres}
\ee
Unfortunately, the amplitude ${\cal A}_1$ has not yet been
measured experimentally.
However, 
if one adopts the estimate
that ${\cal A}_1 = 0.15$ as in (\ref{Avalues}),
then the KTeV asymmetry could be reproduced if 
${\overline{\delta \varphi}} \gappeq 58^o$.

We conclude that, whilst {\it a priori} it may seem very unlikely that the
KTeV asymmetry
could be due to CPT violation, we are unable to exclude rigorously this
possibility at the present time. We hope that future measurements of this
and related decay modes will soon be able to settle this issue.

\section*{Acknowledgements}

We thank members of the CPLEAR and KTeV Collaborations 
for informative
discussions. The work of N.E.M. is partially supported by a 
United Kingdom P.P.A.R.C. Advanced Fellowship.

\end{document}